\documentclass[10pt, a4paper]{article}

\usepackage[final]{lrec2026} 

\newcommand{\Sref}[1]{\S\ref{#1}}
\newcommand{\Fref}[1]{Figure~\ref{#1}}
\newcommand{\Tref}[1]{Table~\ref{#1}}
\usepackage{booktabs}
\usepackage{CJKutf8}
\usepackage{pgfplots}
\usepackage{pgf-pie}
\usepackage{caption}
\usetikzlibrary{matrix,positioning}
\usepackage{multirow}

\title{CS-YODAS: A Mined Dataset of In-the-Wild Code-Switched Speech}

\name{
\parbox{0.98\textwidth}{
\centering
Brian Yan$^1$, Qingzheng Wang$^1$, Matthew Wiesner$^2$, Anuj Diwan$^3$, Olga Iakovenko$^4$\\[-2pt]
Alexander Polok$^5$, Injy Hamed$^6$, Shuichiro Shimizu$^7$, Iris Emerman\\[-2pt]
Thomas Hain$^4$, David R. Mortensen$^1$, Peter Viechnicki, Shinji Watanabe$^1$
}
}

\address{$^1$Carnegie Mellon University, $^2$Johns Hopkins University, $^3$University of Texas at Austin,\\
$^4$University of Sheffield, $^5$Brno University of Technology, $^6$MBZUAI, $^7$Kyoto University\\
byan@cmu.cs.edu
         }

\abstract{
We present CS-YODAS, a Creative Commons-licensed dataset of in-the-wild code-switched speech mined from multilingual YouTube data.
Code-switching (CS), or the alternation between languages within an utterance or conversation, is common in multilingual settings but remains underrepresented in existing CS speech resources, which are typically small, domain-specific, or artificially constructed. 
Building on the YODAS corpus, we develop a scalable, human-in-the-loop pipeline for identifying and validating naturally occurring code-switching. 
The resulting dataset, which totals 313 hours and spans 7 matrix languages, provides diverse, real-world examples of spontaneous code-switched speech. 
We further analyze the distribution and characteristics of code-switching in the wild, examining language-pair frequencies and switching patterns, and report baseline results for spoken language identification. 
We hope that CS-YODAS will encourage broader and more comprehensive research on code-switched speech.
Dataset link: \url{https://huggingface.co/datasets/byan/cs-yodas}.
 \\ \newline  }

\begin{document}

\maketitleabstract

\section{Introduction}

In the current era of large-scale multilingual speech processing, models such as Whisper \cite{radford2023robust}, MMS \cite{pratap2024scaling}, and OWSM \cite{peng2025owsm} enable automatic speech recognition (ASR) and language identification (LID) across hundreds of languages. These advances are supported by massive collections of speech, providing broad coverage across languages and domains \cite{kahn2020libri, pratap2020mls, ardila2020common, wang2021voxpopuli, li2023yodas}. 

Yet despite this progress, most large-scale speech resources remain \textit{monolingual} by design: utterances are assigned a single language label, and data mining pipelines are typically optimized to exclude mixed-language content. 
Consequently, a core feature of multilingual communication, \textit{code-switching}, or the alternation between languages within an utterance or conversation, is largely absent from these corpora.

Collecting code-switched speech at scale presents unique challenges. Unlike monolingual data, it requires bilingual or multilingual speakers who naturally alternate between languages, often in informal or situational contexts. Existing code-switched datasets are therefore limited in scope: they tend to be small, domain-specific, or based on elicited rather than spontaneous speech \cite{lyu2010seame, deuchar2014building, van-der-westhuizen-niesler-2018-first, chowdhury21_interspeech, diwan21_interspeech, hamed2022arzen}. While synthetic approaches have been proposed to fill this gap, they often fail to capture the spontaneity, prosody, and sociolinguistic nuances of naturally occurring multilingual speech \cite{hussein2024speech, yan2025cs}.

In this work, we introduce CS-YODAS, a dataset of naturally occurring, in-the-wild code-switched speech mined from multilingual YouTube recordings.
Our core motivation is based on the premise that better, more natural code-switched speech data will beget better, more natural code-switched speech systems.
We envision that this dataset can be used as a comparison point for synthetically generated code-switching, helping practitioners identify avenues to improve the naturalness of their systems, or as training data for spoken LID systems, improving model robustness beyond what existing domain-specific and synthetically generated data can attain. 


Our contributions are summarized as follows:
\begin{itemize}
    \item A scalable, human-in-the-loop data mining procedure which enables high-precision code-switched detection on in-the-wild data
    \item The mined code-switched speech dataset, which consists of 313 hours of transcribed speech across 7 matrix languages
    \item Empirical characterizations of code-switching in the wild, along with baseline evaluations for CS-aware LID systems
\end{itemize}
Identifying code-switching in large, web-mined corpora presents a major challenge: spontaneous code-switched speech is rare and difficult to detect automatically. 
Off-the-shelf LID tools struggle with identifying multiple languages within utterances, particularly for noisy in-the-wild data, leading to high false-positive rates. 
To address this, we design a scalable, human-in-the-loop pipeline that iteratively refines Large Language Model (LLM)-based code-switched detection using human validation. 
Human feedback is used to guide the LLM via in-context learning, substantially improving the precision of detected code-switched segments while maintaining scalability across web-scale speech data.

The resulting CS-YODAS dataset reflects a wide range of real-world contexts, including conversational, entertainment, and educational domains, capturing both spontaneous and semi-scripted multilingual interactions.
This dataset aims to facilitate future research into both the mechanisms and modeling of natural code-switched speech.



\section{Dataset Construction}
\label{sec:mining}

\subsection{Source Corpus and Motivation}

CS-YODAS is derived from the YODAS corpus \cite{li2023yodas}, a large-scale multilingual speech resource collected from publicly available YouTube content -- by implication, CS-YODAS is under the Creative Commons license.
Specifically, we use the cleaned YODAS from the OWSM v4 project \cite{peng2025owsm} which consists of 166k hours of audio spanning 75 languages, with speech segments paired to automatic transcripts.
While the original corpus organizes data by a single language label, preliminary inspection revealed the presence of code-switching, motivating a systematic effort to identify and extract these examples.

\subsection{Overview}

\begin{figure}[h]
\centering
\includegraphics[width=.75\linewidth]{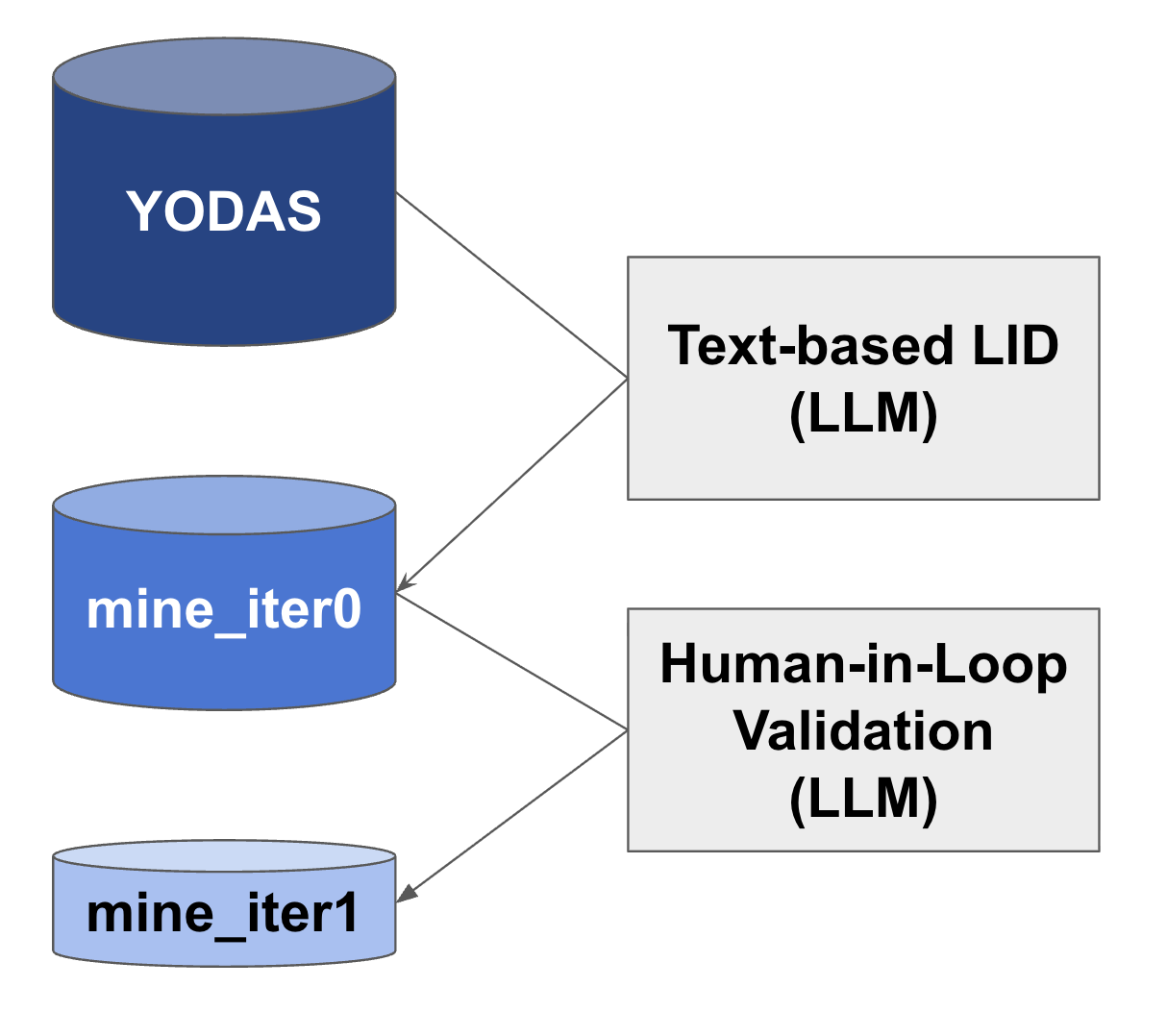}
\caption{Data mining pipeline: we use human feedback on 700 \textit{mine\_iter0} utterances as in-context examples while prompting the LLM to validate all of \textit{mine\_iter0}, thus producing \textit{mine\_iter1}.}
\label{fig:pipeline}
\end{figure}

As shown in \Fref{fig:pipeline}, we apply a LLM-based pipeline to identify code-switched segments: first a text-based LID stage produces a set of candidate segments (\textit{mine\_iter0}) and then a human-in-the-loop validation stage produces the final CS-YODAS set (\textit{mine\_iter1}).
In the remainder of this section, we describe each stage of the pipeline in detail.
We also provide example prompts in \Sref{sec:prompts}.

\subsection{Mining of Candidate Segments}

Each transcript line from YODAS is passed to a multilingual LLM\footnote{\url{https://huggingface.co/Qwen/Qwen3-14B}} which is prompted to infer the set of languages present in the text. 
The model outputs both a primary language (the dominant or matrix language) and a list of all other detected languages. 
Segments where two or more distinct languages are detected are retained as candidate code-switched utterances.
This approach leverages the LLM’s contextual reasoning and language knowledge to capture switching phenomena that may not be easily identified through token-level LID, such as transliteration, named entities, or embedded phrases.

\begin{table}[t]
  \centering
    \resizebox {\linewidth} {!} {
\begin{tabularx}{\linewidth}{lX}
\toprule
Types & Examples \\
\midrule
Transcript Errors & \textcolor{red}{param}\begin{CJK}{UTF8}{min}ことないんですけど今回ね\end{CJK} \\
\midrule
Proper Nouns & \begin{CJK}{UTF8}{mj}어려운 \textcolor{red}{영어를} 참 친엄마 생각하여\end{CJK} \\
\midrule
Cognates & \textcolor{red}{moment} là \textcolor{red}{pour} la \textcolor{red}{phase} 2 seulement deux\\
\bottomrule
\end{tabularx}
}
    \caption{Examples of common non-CS distractors: words in \textcolor{red}{red} caused text-based LID to incorrectly predict the presence of English.}
    \label{tab:distractors}
\end{table}

However, the in-the-wild nature of the source YODAS data introduces many \textit{distractors}: noisy or imperfect transcripts can falsely mix languages, and borrowed words or named entities often resemble code-switching without representing genuine language alternation.
\Tref{tab:distractors} presents examples of the three most common types: transcription errors, proper nouns, and cognates.

\subsection{Human-in-the-Loop Validation}
\label{sec:human}

To improve precision, we introduce a human-in-the-loop validation procedure.
We sample 100 candidate segments across 7 matrix languages for human validation: Arabic, Chinese Mandarin, Czech, French, Hindi, Japanese, Russian.
For each segment, annotators are asked 5 questions:
\begin{enumerate}
    \item Is the transcript correct?
    \item Does the segment contain <lang1>?
    \item Is <lang1> the matrix language?
    \item Does the segment contain <lang2>?
    \item Are all <lang2> words proper nouns?
\end{enumerate}
Each of these must be answered with ``Yes'', ``No'', or ``I can't tell''. Additionally, we solicit comments for each segment where any of Q1-4 were not answered with "Yes". 

This human feedback is then utilized via in-context learning when prompting the LLM to generate responses to the same 5 questions for all candidate segments (see \Sref{sec:prompts} for example prompts). 
Note: we also allow the LLM to generate comments.
We then filter the candidate segments using the following criteria: Q1-4 = "Yes" and Q5 = "No".
The resulting set is taken as the final CS-YODAS set (\textit{mine\_iter1}).

\subsection{Evaluating the Pipeline}
\label{sec:evaluating}

We conducted human evaluation on 200 candidate segments drawn from the \textit{mine\_iter0} set for each of the seven aforementioned matrix languages.
Of these, 100 samples were used as in-context examples for the LLM-based validation round, while the remaining 100 were passed through the full data mining pipeline, with some ultimately appearing in the final validated set, \textit{mine\_iter1}.

\begin{table}[t]
  \centering
\begin{tabular}{lcc}
\toprule
Set & Samples Evaluated & Precision\\
\midrule
mine\_iter0 & 700 & 18.0\% \\
mine\_iter1 & 120 & 70.0\%\\
\bottomrule
\end{tabular}
    \caption{Precision of samples selected from \textit{mine\_iter0} vs \textit{mine\_iter1}, measured via agreement with human evaluations.}
    \label{tab:precision}
\end{table}

As shown in \Tref{tab:precision}, 700 candidate segments from \textit{mine\_iter0} were evaluated and only 18\% were confirmed to be true instances of code-switching.
Following the human-in-the-loop validation, the resulting \textit{mine\_iter1} set reached 70\% precision across 120 segments identified as code-switched.

\begin{table}[t]
  \centering
    \resizebox {\linewidth} {!} {
\begin{tabular}{lcccccc}
\toprule
Lang & TP & FP & FN & TN & Pre. (\%) & Rec. (\%) \\
\midrule
ara & 16 & 13 & 20 & 51 & 55.2 & 44.4 \\
ces & 1 & 1 & 1 & 97 & 50.0 & 50.0 \\
cmn & 20 & 6 & 7 & 67 & 76.9 & 74.1 \\
fra & 9 & 6 & 13 & 72 & 60.0 & 40.9 \\
hin & 34 & 7 & 8 & 51 & 82.9 & 81.9 \\
jpn & 2 & 3 & 10 & 50 & 40.0 & 16.7 \\
rus & 2 & 0 & 28 & 70 & 100.0 & 6.7 \\
\midrule
all & 84 & 36 & 87 & 458 & 70.0 & 49.1\\
\bottomrule
\end{tabular}
}
    \caption{Full confusion matrix breakdown of the human-in-the-loop stage of the data mining pipeline. TP = True Positive, FP = False Positive, FN = False Negative, Pre. = Precision, Rec. = Recall.}
    \label{tab:precision_v_recall}
\end{table}

\Tref{tab:precision_v_recall} further breaks down the effect of human-in-the-loop validation, providing the full confusion matrix across the 7 matrix languages.
Hindi, Chinese, and Arabic yielded the most true positive code-switched samples at precision rates of 82.9\%, 76.9\%, and 55.2\% respectively.
Japanese and Russian yielded very few true positives; we found that samples with only proper nouns in the embedded language made up a large portion of the false negatives, indicating some annotator disagreement with our filtering methodology (namely requiring Q5 = ``No'').
These results highlight variation in the reliability of automatic code-switch detection across matrix languages, underscoring the importance of gathering language-specific human feedback.





\subsection{Capturing Code-Switching Context}

\begin{table}[t]
  \centering
    \resizebox {\linewidth} {!} {
\begin{tabular}{lccc}
\toprule
Lang & CS Duration & Total Duration & CS Rate (\%) \\
\midrule
ara & 0.6 & 2.8 & 21.4 \\
ces & 0.0 & 0.3 & 13.7 \\
cmn & 0.5 & 4.1 & 12.1 \\
fra & 28.5 & 273.7 & 10.4 \\
hin & 6.3 & 21.2 & 29.7 \\
jpn & 0.3 & 2.0 & 15.0 \\
rus & 1.1 & 8.9 & 12.4 \\
\midrule
all & 37.3 & 312.7 & 11.9 \\
\bottomrule
\end{tabular}
}
    \caption{Durations (hours) per matrix language: "CS Duration" refers to intra-sentential CS segments while ``Total Duration'' refers to contextual chunks (CS segment + 15 sec of left/right pad). CS Rate = CS Duration / Total Duration.}
    \label{tab:context}
\end{table}

With CS-YODAS, we aim to facilitate the study of when and where code-switching arises in real-world communication contexts.
To this end, we construct CS-YODAS in a contextual manner as follows.
As described in the previous sections, we first mine a set of target segments with intra-sentential code-switching -- these are typically short (<5 sec) utterances.
We then extract the surrounding contexts (15 sec before and after) for these target segments and concatenate, producing a contextual chunk.
In the case that multiple target segments are within 30 seconds of each other, we simply merge them into one contiguous contextual chunk.

As shown in \Tref{tab:context}, the total duration of contextual audio for the seven matrix languages is 312.7 hours.
Of this, 37.3 hours (11.9\%) correspond to intra-sententially code-switched speech.
Certain languages, such as Arabic and Hindi, exhibit a higher code-switching rate.







\subsection{Dataset Statistics}

\begin{table}[h]
  \centering
    \resizebox {\linewidth} {!} {
\setlength{\tabcolsep}{2pt}
\begin{tabular}{lccccccc}
\toprule
Dataset & Matrix Langs & Domain & Speech Type & Hours \\
\midrule
CS-FLEURS & 52 & Wikipedia & Read/Synthetic & 294 \\
- read-set & 14 & Wikipedia & Read & 17 \\
- xtts-set & 16 & Wikipedia & Synthetic & 221 \\
- mms-set & 45 & Wikipedia & Synthetic & 56\\
\midrule
CS-YODAS & 7 & Youtube & Spontaneous & 313 \\
\bottomrule
\end{tabular}
}
    \caption{Summary of CS-YODAS vs. CS-FLEURS.}
    \label{tab:summary}
\end{table}

Summary statistics are presented in \Tref{tab:summary}.
The final CS-YODAS release comprises 313 hours of speech across 7 matrix languages. 

Compared to previous collections, CS-YODAS uniquely offers spontaneous code-switched speech across a wide range of languages.
Earlier efforts to expand the language coverage of code-switched corpora have largely relied on synthetic generation. For example, CS-FLEURS \cite{yan2025cs} was created by first generating code-switched Wikipedia text and then recording it as read speech or synthesizing it using TTS.
While such methods are valuable for scaling data collection, they fall short in capturing the natural dynamics of code-switching: for instance, when and why speakers alternate between languages in real contexts.






\section{Dataset Analyses}

\subsection{Overview}

The primary motivation of our dataset analysis is to augment the dataset with additional metadata, such as language roles, parts-of-speech, and text or audio domains.
In this section, we describe our methodology for generating metadata and summarize the metadata distributions.

Please be aware that the following analyses are based on data which we estimate to be of 70\% precision (as described in section \Sref{sec:evaluating}). 
Please also refer to \Sref{sec:limitations} for further discussion of the limitations of this dataset.



\subsection{Language Distributions}

\begin{figure}[h]
\centering
\resizebox {\linewidth} {!} {
\begin{tikzpicture}
\begin{axis}[
    ybar,
    bar width=30pt,
    width=\textwidth,
    height=0.5\textwidth,
    enlargelimits=0.1,
    legend style={at={(0.5,-0.15)}, anchor=north, legend columns=-1},
    ymin=0,
    ymax=8,
    enlarge y limits={abs=0},
    ylabel={\LARGE Yield ($\%$)},
    ylabel style={font=\LARGE},
    symbolic x coords={ara, ces, cmn, fra, hin, jpn, rus},
    xtick=data,
    tick label style={font=\LARGE},
    nodes near coords,
    nodes near coords align={vertical},
    nodes near coords style={font=\LARGE, /pgf/number format/fixed},
]

\addplot[fill=lightgray] coordinates {
    (ara,1.64)
    (ces,0.02)
    (cmn,0.52)
    (fra,0.66)
    (hin,6.87)
    (jpn,0.03)
    (rus,0.01)
};

\end{axis}
\end{tikzpicture}

}
\vspace{-2em}
\caption{Yield (\%) of segments identified as code-switched out of the total number of segments across matrix languages.}
\label{fig:yield}
\end{figure}
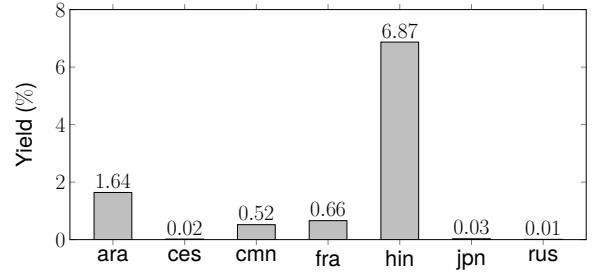


First, let's address the obvious question: how often did code-switching occur in-the-wild?
Given that our method prioritizes precision over recall we cannot directly answer this question.
Instead, we try to provide a lower-bound estimate by reporting the data mining yield rate (i.e. the number of segments identified as code-switched out of the total number of source segments) -- we expect that the actual rate of code-switching in-the-wild is somewhat higher than the yield rate.
As shown in \Fref{fig:yield}, the yield varied greatly across matrix languages, ranging from 0.01\% for Russian to 6.87\% Hindi.
Arabic (1.64\%), French (0.66\%), and Chinese (0.52\%) also exhibited moderate yield rates.

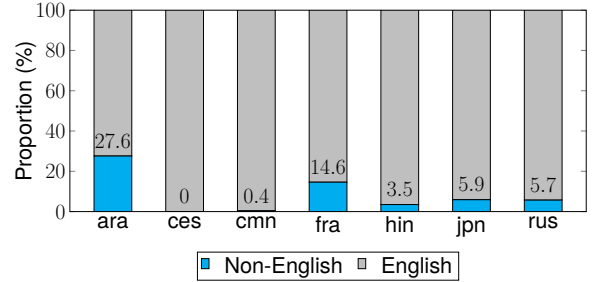
\begin{figure}[t]
\centering
\resizebox{\linewidth}{!}{
\begin{tikzpicture}
\begin{axis}[
    ybar stacked,
    bar width=30pt,
    width=\textwidth,
    height=0.45\textwidth,
    ymin=0,
    ymax=100,
    ylabel={\LARGE Proportion (\%)},
    ylabel style={font=\LARGE},
    symbolic x coords={ara, ces, cmn, fra, hin, jpn, rus},
    xtick=data,
    tick label style={font=\LARGE},
    legend style={
        at={(0.5,-0.2)},
        anchor=north,
        legend columns=-1,
        font=\Large,
        column sep=1em,
    },
]
\addplot[fill=cyan,
    nodes near coords,
    nodes near coords align={vertical},
    every node near coord/.append style={
        font=\LARGE,
        yshift=3pt,
        color=black,
    }] 
coordinates {
    (ara,27.6)
    (ces,0.0)
    (cmn,0.4)
    (fra,14.6)
    (hin,3.5)
    (jpn,5.9)
    (rus,5.7)
};
\addplot[fill=lightgray]
coordinates {
    (ara,72.4)
    (ces,100)
    (cmn,99.6)
    (fra,85.4)
    (hin,96.5)
    (jpn,94.1)
    (rus,94.3)
};
\legend{\LARGE Non-English, \LARGE English}
\end{axis}
\end{tikzpicture}
}
\caption{Proportions (\%) of examples with English vs. Non-English as the embedded language across matrix languages.}
\label{fig:embedded}
\end{figure}

Next, let's examine the distribution of embedded languages.
English is by far the most common embedded language, accounting for 85.6\% of the data.
\Fref{fig:embedded} shows the English vs. Non-English embedded language breakdown across the 7 matrix languages.
Arabic (27.6\%) and French (14.6\%) exhibit the highest rates of code-switching with non-English languages.
For Arabic, Egyptian Arabic accounted for the highest proportion of non-English code-switching. 
For French, Arabic accounted for the highest rate of non-English code-switching.

\subsection{Lexical Distributions}

\begin{figure}[h]
\centering
\resizebox{\linewidth}{!}{
\begin{tikzpicture}
\begin{axis}[
    xbar stacked,
    bar width=30pt,
    width=\textwidth,
    height=0.4\textwidth,
    xmin=0, xmax=100,
    xlabel={Proportion (\%)},
    xlabel style={font=\LARGE},
    symbolic y coords={CS-YODAS, CS-FLEURS},
    ytick=data,
    nodes near coords,
    enlarge y limits=0.6,
    tick label style={font=\LARGE},
    legend style={
        at={(0.5,-0.35)},
        anchor=north,
        legend columns=2,
        font=\LARGE,
        column sep=2em
    },
    point meta=rawx,
    nodes near coords align={center},
    every node near coord/.append style={
        font=\LARGE,
        color=black,
        anchor=east,
    },
]

\addplot[fill=violet!30] coordinates {(48.2,CS-YODAS) (46.0,CS-FLEURS)};
\addplot[fill=lightgray] coordinates {(40.2,CS-YODAS) (36.7,CS-FLEURS)};
\addplot[fill=red!30] coordinates {(8.4,CS-YODAS) (16.5,CS-FLEURS)};
\addplot[fill=cyan!40, every node near coord/.append style={
        font=\LARGE,
        color=black,
        anchor=west
    },] coordinates {(3.2,CS-YODAS) (0.8,CS-FLEURS)};

\legend{Content Words, Proper Nouns, Function Words, Discourse Markers}

\end{axis}
\end{tikzpicture}
}
\vspace{-2em}
\caption{Proportions (\%) of embedded English words across 4 categories: content words, proper nouns, function words, and discourse markers.}
\label{fig:pos}
\end{figure}
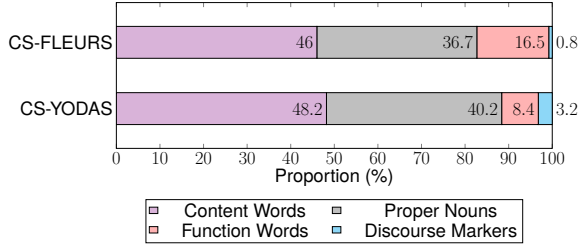

Next, we examine the lexical distribution of the embedded English words to better understand the conversational purpose of code-switching. 
Our analysis focuses on English words to ensure reliable part-of-speech (POS) tagging. 
We first extract all English words and run POS tagging using spaCy. 
Each word is then assigned to one of four categories: \textit{content words}, \textit{function words}, \textit{proper nouns}, and \textit{discourse markers}. 
Content words include nouns, verbs, adjectives, and adverbs. 
Proper nouns are treated as a separate category.
Function words include determiners, adpositions, conjunctions, particles, and auxiliaries. 
Finally, discourse markers include interjections such as ``yeah'', ``like'', ``well'', etc.

\Fref{fig:pos} plots the proportions of the four lexical categories for both CS-YODAS and CS-FLEURS; the denominator here is the total count of embedded English words.
Both sets contain mostly content words and proper nouns among their embedded English words. 
However, CS-YODAS shows a much smaller proportion of function words, nearly half that of CS-FLEURS -- this suggests that the lexical pattern of the synthesized CS-FLEURS text may be unnatural. This finding aligns with linguistic theory, which suggests that function words should be least commonly used in the embedded language~\cite{myers_scotton93}.
Finally, CS-YODAS includes some discourse markers, which are typical in conversational speech, whereas CS-FLEURS has none, consistent with its synthetic nature.





\subsection{Domain Distributions}

\begin{figure}[h]
\centering
\includegraphics[width=\linewidth]{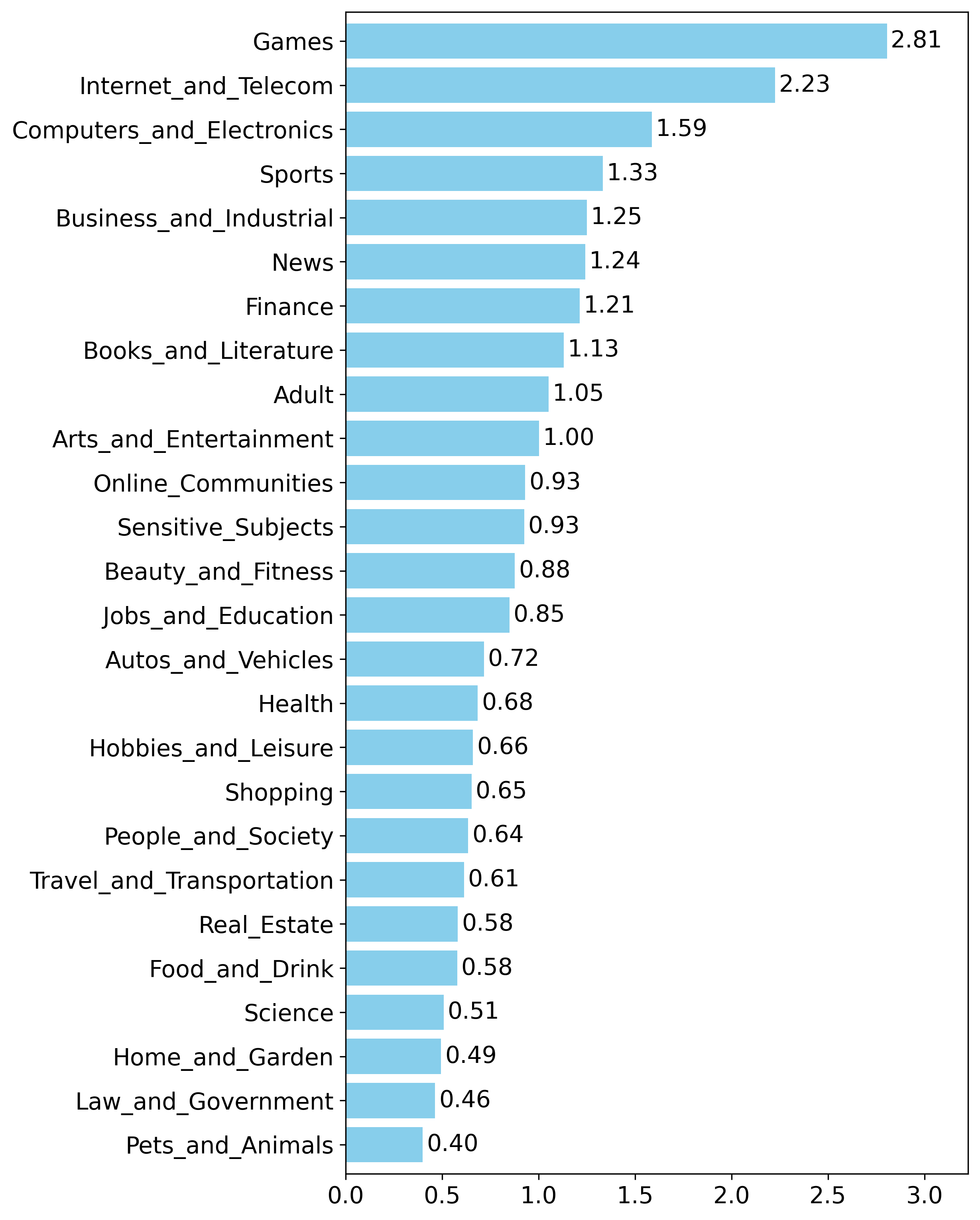}
\caption{CS likelihood by domain (CS-YODAS proportion / YODAS proportion): values greater/less than 1 suggest higher/lower CS likelihood.}
\label{fig:ratio}
\end{figure}

To further characterize the dataset, we obtain topical domain labels by feeding transcripts to a multilingual text-based domain classification model.\footnote{\url{https://huggingface.co/nvidia/multilingual-domain-classifier}}
We then use these domain labels to estimate where code-switching tends to occur. 
To do so, we compare the proportional representation of each domain in CS-YODAS to that in the original YODAS corpus. 
For each domain, we compute the ratio between its relative frequency in CS-YODAS and its relative frequency in YODAS. 
A value greater than one indicates that the domain is over-represented in CS-YODAS relative to the source YODAS data, while values below one indicate under-representation. 
This approach normalizes for overall corpus composition and highlights domains where code-switching occurs disproportionately often.

As shown in \Fref{fig:ratio}, 
informal and tech-oriented domains such as \textit{Games}, \textit{Internet and Telecom}, and \textit{Computers and Electronics} show the strongest over-representation, suggesting that code-switching frequently arises in conversational, online, and technology-related contexts. 
Moderately elevated ratios are also observed in \textit{Sports}, \textit{Business}, and \textit{News}, where English terms or expressions are often borrowed. 
In contrast, more formal or topic-specific domains, such as \textit{Law and Government}, \textit{Health}, \textit{Science}, and \textit{People and Society}, show lower ratios, indicating that code-switching is less common in structured or institutional settings. 

\section{Baseline Experiments}

\begin{table*}[h]
  \centering
    \resizebox{\linewidth}{!}{
\setlength{\tabcolsep}{4pt}
\begin{tabular}{lcccccccccc}
  \toprule
  \multirow{4}{*}{Train Set} & \multirow{4}{*}{FLEURS} & \multicolumn{8}{c}{CS-FLEURS} \\
  \cmidrule(lr){3-10}
   &  & \multirow{2.5}{*}{XTTS1} & \multirow{2.5}{*}{MMS} & \multicolumn{6}{c}{READ} \\
  \cmidrule(lr){5-10}
   &  &  & & ara-eng & cmn-eng & fra-eng & hin-eng & jpn-eng & rus-eng \\
  \midrule
  w/o CS-YODAS & 97.4 & 99.7 & 99.9 & 0  & 0 & 0 & 0 & 0 & 0 \\
  w/ CS-YODAS & 96.3 & 99.5 & 99.8  & 0 & 0.3 & 51.1 & 19.3 & 0 & 0 \\
  \bottomrule
\end{tabular}
}
    \caption{Comparison of spoken LID models with and without CS-YODAS training data. Accuracy (\%) is reported on FLEURS and CS-FLEURS.}
    \label{tab:slid}
\end{table*}

\subsection{Overview}

Modern multilingual speech recognition and translation systems rely on accurate spoken LID both for curating large-scale training datasets~\cite{valk2021voxlingua107, peng2025owsm} and for routing input to the language-specific modules~\cite{radford2023robust, pratap2024scaling}.
However, current spoken LID systems~\cite{jia23b_interspeech, wang2025geolocation} are almost exclusively trained under the assumption that utterances are purely monolingual, ignoring the possibility that a single utterance may contain multiple languages. 
Prior work has shown that this monolingual bias greatly limits the quality of downstream speech processing for code-switched speech \cite{peng2023prompting, yan2025cs}.

This limitation reflects a broader gap in LID research.
\citet{burchell2024code} show that even in the text domain, utterance-level code-switched LID remains challenging.
In the spoken domain, prior LID studies~\cite{rangan2020exploiting, li2023simple} are limited to only a few language pairs.
We argue that this gap primarily stems from a data bottleneck: large-scale multilingual speech corpora with natural code-switching and reliable language labels have been lacking.

In this work, we leverage CS-YODAS alongside the monolingual FLEURS dataset and the synthetic code-switched CS-FLEURS dataset to establish baseline LID systems that explicitly account for both monolingual and code-switched utterances.
Our findings show that synthetic code-switched training data alone is insufficient to support robust generalization, and that exposure to spontaneous, naturally occurring code-switching is essential.
These results highlight the importance and feasibility of curating large-scale spontaneous code-switched resources to enable more realistic spoken LID in the wild.

\subsection{Experimental Setup}

We compare two training data configurations:
(1) simply combining the FLEURS~\cite{conneau2023fleurs} training set with the XTTS-generated training data from CS-FLEURS and (2) further including CS-YODAS data.\footnote{We only use the examples with English as the embedded language for this experiment, and exclude ces due to its limited data.}
Both configurations contain 102 monolingual languages and 16 code-switched language pairs. The 16 pairs originate from CS-FLEURS, with 6 of them further supplemented by data from CS-YODAS.
We refer to these settings as ``w/o CS-YODAS'' and ``w/ CS-YODAS'' respectively.

Our LID models are based on self-supervised representations, with MMS~\cite{pratap2024scaling} as the upstream encoder and ECAPA-TDNN~\cite{desplanques20_interspeech} as the downstream embedding extractor.
Classification is performed using the AAMSoftmax loss~\cite{deng2019arcface} with the sub-center enhancement~\cite{zhao2021speakin}, following the ESPnet-SPK implementation~\cite{jung2024espnet}.
For code-switched utterances, each language pair is treated as a distinct class (i.e. ``cmn'' and ``cmn-eng'' are distinct).
Training hyperparameters and optimization settings are kept consistent with the VoxLingua107-only setup described in~\citet{wang2025geolocation}.
The final models are selected based on the best accuracy on the FLEURS dev set and CS-YODAS XTTS1 test set.
Models are evaluated on FLEURS and CS-FLEURS test sets.

\subsection{Results and Discussion}

As shown in Table~\ref{tab:slid}, the model ``w/o CS-YODAS'' performs well on in-domain settings, including FLEURS and the CS-FLEURS XTTS1, as well as the out-of-domain synthetic CS-FLEURS MMS.
However, the model fails across all languages in CS-FLEURS READ; the model is unable to distinguish code-switched utterances.
We posit that this is due to a domain mismatch resulting from a lack of natural code-switched training data.

After incorporating the CS-YODAS training data, performance improves on the fra–eng (0\% to 51.1\%) and hin–eng (0\% to 19.9\%) subsets of the CS-FLEURS READ.
This \textit{zero-to-one} improvement demonstrates that exposure to in-the-wild code-switched speech enables models to begin generalizing to natural switching patterns beyond what synthetic data can support.

\begin{figure}[h]
\centering
\resizebox{\linewidth}{!}{
\begin{tikzpicture}
\begin{axis}[
    width=\linewidth,
    height=0.5\linewidth,
    xlabel={\small Duration (hours)},
    ylabel={\small Accuracy (\%)},
    y label style={at={(axis description cs:0.12,0.5)},anchor=south},
    xmin=0, xmax=30,
    ymin=0, ymax=55,
    xtick={0,5,10,15,20,25,30},
    ytick={0,20,40,60},
    grid=major,
    grid style={dashed, gray!30},
    tick label style={font=\small},
    label style={font=\small},
    every axis plot/.append style={thick, mark=*}
]

\addplot[color=blue, mark=*] coordinates {
    (0.01, 0)
    (0.3, 0)
    (0.6, 0)
    (0.5, 0.3)
    (1.1, 0)
    (6.3, 19.3)
    (28.5, 51.1)
};

\end{axis}
\end{tikzpicture}
}
\vspace{-2em}
\caption{Accuracy (\%) on CS-FLEURS READ vs. Duration (hours) of CS-YODAS training data.}
\label{fig:trend}
\end{figure}
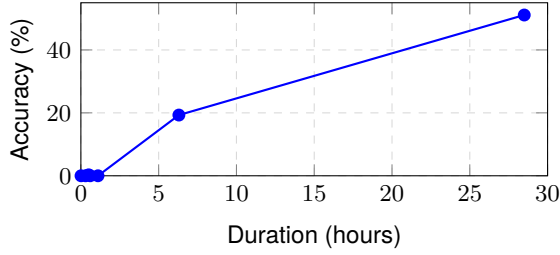

As shown in \Fref{fig:trend}, there is a clear trend between code-switched training data duration and code-switched LID performance: accuracy does not rise past 0 until after 5 hours of code-switched training data is available. 

Our hope is that these baselines demonstrating the impact of in-the-wild code-switched training data motivates future dataset construction efforts, as our belief is that solving the data problem is critical towards building effective code-switched spoken LID systems.

\section{Conclusion}

In summary, our work introduces CS-YODAS, a large-scale dataset of spontaneous, naturally occurring code-switched speech.
Using CS-YODAS along with the existing CS-FLEURS, we are able to show that synthetically generated code-switching has yet to thoroughly mimic the patterns observed in the wild.
Further, we show that synthetic code-switched training data has yet to obviate the need for collecting spontaneous code-switched speech. 
Our hope is that CS-YODAS facilitates future research towards building robust spoken LID systems capable of handling all nuances of  multilingual communication.



\section{Limitations}
\label{sec:limitations}

While CS-YODAS represents a significant step toward large-scale, naturally occurring code-switched speech resources, several limitations remain.

First, the underlying source corpus, YODAS, is based on Creative Commons YouTube content, which inherently biases the dataset toward publicly available and broadcast-style material, such as news, educational talks, and announcements.
While the dataset still captures code-switching in daily, informal contexts, we suspect that the casual setting may be underrepresented.

Second, while we implemented a human-in-the-loop validation pipeline to improve the precision of mined code-switched examples, the scale of the data precludes exhaustive manual verification. 
We also acknowledge other the presence of residual noise from the source corpus, such as transcription and audio segmentation errors.

Finally, reliable evaluation of code-switching requires skilled bilingual annotators. 
This reliance on language-specific expertise is a limiting factor against scaling coverage, and thus this release of CS-YODAS spans only 7 matrix languages. 


\section{Ethics Statement}

This dataset is a mined subset from an already publicly released dataset, so we do not foresee any harm arising from its content.
Annotators were recruited for this project on a volunteer basis and were made aware of our data mining approach and thus agreed to the use of their feedback to refine the dataset.

\section{Example LLM Prompts}
\label{sec:prompts}

\subsection{Overview}

This section provides examples of the prompts used for text-based LID and human-in-the-loop Validation, supplementing the main description of the LLM-based data mining pipeline in \Sref{sec:mining}.

\subsection{Text-based LID}

\begin{table}[h]
   \centering
     \footnotesize
\resizebox {\linewidth} {!} {
\setlength{\tabcolsep}{2.5pt}
\begin{tabular}{p{\linewidth}}
\toprule
(System) You are performing text-based language identification. We are trying to identify code-mixed or code-switched utterances. \\
\\
\textcolor{blue}{(User) Text: ```\begin{CJK}{UTF8}{gbsn}这些 地区 人口 稀少 ， often 不 存在 光 污染 的 问题 ， 你 also 能 欣赏 到 璀 璨 星空 。\end{CJK}```}\\
\\
\textcolor{blue}{For the given text in triple backticks identify ALL languages that appear. There may be only a single language or multiple languages that are code-mixed together. Your final answer should list the languages in order of prevalence.}\\
\textcolor{blue}{Code-mixing, or code-switching, is defined as the alternation of two languages within a single discourse, sentence, or constituent. Double check whether the text contains code-switching by reviewing word-by-word. Do not simply glance at the overall sentence and only write down the dominant language.}\\
\textcolor{blue}{Format your response as a json object.}\\
\\
\textcolor{blue}{(Assistant) \{"languages": ["Chinese", "English"]\}}\\
\\
\textcolor{blue}{\textit{... 3 in-context examples are provided in total ...}}\\
\\
\textcolor{violet}{(User) <Prompt with target segment>}\\
\textcolor{violet}{(Assistant) <Text-based LID output>}\\
\bottomrule
\end{tabular}
}
     \caption{Example prompt used for text-based LID. In-context examples are depicted in \textcolor{blue}{blue} and target segment prompt/response are depicted in \textcolor{violet}{violet}.}
     \label{tab:lid_prompt}
\end{table}

\Tref{tab:lid_prompt} illustrates an example prompt used for the first stage of our data mining pipeline.
We task the LLM with identifying all languages that appear in the segment.
We use only 3 in-context examples in this stage; this choice was made in an attempt to maximize throughput.

\subsection{Human-in-the-Loop Validation}

 \begin{table}[h]
   \centering
     \footnotesize
\resizebox {\linewidth} {!} {
\setlength{\tabcolsep}{2.5pt}
\begin{tabular}{p{\linewidth}}
\toprule
(System) You are performing data validation. We are looking for code-switched speech examples by examining transcript text.\\
\\
\textcolor{blue}{(User) Text: ```\begin{CJK}{UTF8}{gbsn}因此受到他的祝福之下就代表的是权威Authority\end{CJK}```}\\
\\
\textcolor{blue}{For the given text in triple backticks, answer the following 5 questions:}\\
\textcolor{blue}{1. Is the transcript correct?}\\
\textcolor{blue}{2. Does the speech contain Chinese?}\\
\textcolor{blue}{3. Is Chinese the matrix language?}\\
\textcolor{blue}{4. Does the speech contain English?}\\
\textcolor{blue}{5. Are all English words proper nouns?}\\
\\
\textcolor{blue}{(Assistant) \{"Q1": "Yes", "Q2": "Yes", "Q3": "Yes", "Q4": "Yes", "Q5": "No", "Comments": ""\}}\\
\\
\textcolor{blue}{... \textit{100 in-context examples are provided in total }...}\\
\\
\textcolor{violet}{(User) <Prompt with target segment>}\\
\textcolor{violet}{(Assistant) <Human-in-the-loop Validation output>}\\
\bottomrule
\end{tabular}
}
     \caption{Example prompt used for human-in-the-loop validation. In-context examples are depicted in \textcolor{blue}{blue} and target segment prompt/response are depicted in \textcolor{violet}{violet}.}
     \label{tab:val_prompt}
 \end{table}

\Tref{tab:val_prompt} illustrates an example prompt used for the second stage of our data mining pipeline.
We task the LLM with replicating the human feedback pattern.
We use 100 in-context examples in this stage, sourced from the human feedback on same-language samples; this choice was made in an attempt to maximize accuracy.

\section{References}\label{sec:reference}

\bibliographystyle{lrec2026-natbib}
\bibliography{mybib}

@Inproceedings{radford2023robust,
  title = "Robust speech recognition via large-scale weak supervision",
  author = "Radford, Alec and Kim, Jong Wook and Xu, Tao and Brockman, Greg and McLeavey, Christine and Sutskever, Ilya",
  booktitle = "Proc. ICML",
  pages = "28492--28518",
  year = 2023
}

@Article{pratap2024scaling,
  title = "Scaling speech technology to 1,000+ languages",
  author = "Pratap, Vineel and Tjandra, Andros and Shi, Bowen and Tomasello, Paden and Babu, Arun and Kundu, Sayani and Elkahky, Ali and Ni, Zhaoheng and Vyas, Apoorv and Fazel-Zarandi, Maryam and others",
  journal = "Journal of Machine Learning Research",
  pages = "1--52",
  year = 2024
}

@Article{peng2025owsm,
  title = "{OWSM} v4: Improving open {Whisper}-style speech models via data scaling and cleaning",
  author = "Peng, Yifan and Muhammad, Shakeel and Sudo, Yui and Chen, William and Tian, Jinchuan and Lin, Chyi-Jiunn and Watanabe, Shinji",
  journal = "Proc. Interspeech",
  pages     = {2225--2229},
  year = 2025
}

@Inproceedings{li2023yodas,
  title = "{YODAS}: {YouTube}-oriented dataset for audio and speech",
  author = "Li, Xinjian and Takamichi, Shinnosuke and Saeki, Takaaki and Chen, William and Shiota, Sayaka and Watanabe, Shinji",
  booktitle = "Proc. IEEE ASRU",
  pages = "1--8",
  year = 2023
}

@Inproceedings{ardila2020common,
  title = "{Common Voice}: A Massively-Multilingual Speech Corpus",
  author = "Ardila, Rosana and Branson, Megan and Davis, Kelly and Kohler, Michael and Meyer, Josh and Henretty, Michael and Morais, Reuben and Saunders, Lindsay and Tyers, Francis and Weber, Gregor",
  booktitle = "Proc. LREC",
  pages = "4218--4222",
  year = 2020
}

@Inproceedings{kahn2020libri,
  title = "{Libri-Light}: A benchmark for {ASR} with limited or no supervision",
  author = "Kahn, Jacob and Riviere, Morgane and Zheng, Weiyi and Kharitonov, Evgeny and Xu, Qiantong and Mazar{\'e}, Pierre-Emmanuel and Karadayi, Julien and Liptchinsky, Vitaliy and Collobert, Ronan and Fuegen, Christian and others",
  booktitle = "Proc. ICASSP",
  pages = "7669--7673",
  year = 2020
}

@Inproceedings{pratap2020mls,
  title = "{MLS}: A Large-Scale Multilingual Dataset for Speech Research",
  author = "Vineel Pratap and Qiantong Xu and Anuroop Sriram and Gabriel Synnaeve and Ronan Collobert",
  booktitle = "Proc. Interspeech",
  pages = "2757--2761",
  year = 2020
}

@Inproceedings{wang2021voxpopuli,
  title = "{VoxPopuli}: A Large-Scale Multilingual Speech Corpus for Representation Learning, Semi-Supervised Learning and Interpretation",
  author = "Wang, Changhan and Riviere, Morgane and Lee, Ann and Wu, Anne and Talnikar, Chaitanya and Haziza, Daniel and Williamson, Mary and Pino, Juan and Dupoux, Emmanuel",
  booktitle = "Proc. ACL-IJCNLP",
  year = 2021
}

@Inproceedings{yan2025cs,
  title = "{CS-FLEURS}: A Massively Multilingual and Code-Switched Speech Dataset",
  author = "Yan, Brian and Hamed, Injy and Shimizu, Shuichiro and Lodagala, Vasista Sai and Chen, William and Iakovenko, Olga and Talafha, Bashar and Hussein, Amir and Polok, Alexander and Chang, Kalvin and others",
  booktitle = "Proc. Interspeech",
  pages = "743--747",
  year = 2025
}

@Inproceedings{conneau2023fleurs,
  title = "{FLEURS}: Few-shot learning evaluation of universal representations of speech",
  author = "Conneau, Alexis and Ma, Min and Khanuja, Simran and Zhang, Yu and Axelrod, Vera and Dalmia, Siddharth and Riesa, Jason and Rivera, Clara and Bapna, Ankur",
  booktitle = "Proc. IEEE SLT",
  pages = "798--805",
  year = 2023
}

@Article{deuchar2014building,
  title = "Building bilingual corpora",
  author = "Deuchar, Margaret and Davies, Peredur and Herring, Jon and Couto, M Carmen Parafita and Carter, Diana",
  journal = "Advances in the Study of Bilingualism",
  pages = "93--111",
  year = 2014
}

@Inproceedings{lyu2010seame,
  title = "{SEAME}: a Mandarin-English code-switching speech corpus in south-east asia",
  author = "Lyu, Dau-Cheng and Tan, Tien Ping and Chng, Engsiong and Li, Haizhou",
  booktitle = "Proc. Interspeech",
  pages = "1986--1989",
  year = 2010
}

@Inproceedings{hamed2022arzen,
  title = "{ArzEn-ST}: A Three-way Speech Translation Corpus for Code-Switched {E}gyptian {A}rabic-{E}nglish",
  author = "Hamed, Injy and Habash, Nizar and Abdennadher, Slim and Vu, Ngoc Thang",
  booktitle = "Proc. WANLP",
  pages = "119--130",
  year = 2022
}

@Inproceedings{diwan21_interspeech,
  title = "{MUCS} 2021: Multilingual and Code-Switching {ASR} Challenges for Low Resource {Indian} Languages",
  author = "Anuj Diwan and Rakesh Vaideeswaran and Sanket Shah and Ankita Singh and Srinivasa Raghavan and Shreya Khare and Vinit Unni and Saurabh Vyas and Akash Rajpuria and Chiranjeevi Yarra and Ashish Mittal and Prasanta Kumar Ghosh and Preethi Jyothi and Kalika Bali and Vivek Seshadri and Sunayana Sitaram and Samarth Bharadwaj and Jai Nanavati and Raoul Nanavati and Karthik Sankaranarayanan",
  booktitle = "Proc. Interspeech",
  pages = "2446--2450",
  year = 2021
}

@Inproceedings{van-der-westhuizen-niesler-2018-first,
  title = "A First South {African} Corpus of Multilingual Code-switched Soap Opera Speech",
  author = "van der Westhuizen, Ewald and Niesler, Thomas",
  booktitle = "Proc. LREC",
  year = 2018
}

@Inproceedings{chowdhury21_interspeech,
  title = "Towards One Model to Rule All: Multilingual Strategy for Dialectal Code-Switching {Arabic} {ASR}",
  author = "Shammur Absar Chowdhury and Amir Hussein and Ahmed Abdelali and Ahmed Ali",
  booktitle = "Proc. Interspeech",
  pages = "2466--2470",
  year = 2021
}

@Inproceedings{hussein2024speech,
  title = "Speech collage: code-switched audio generation by collaging monolingual corpora",
  author = "Hussein, Amir and Zeinali, Dorsa and Klejch, Ond{\v{r}}ej and Wiesner, Matthew and Yan, Brian and Chowdhury, Shammur and Ali, Ahmed and Watanabe, Shinji and Khudanpur, Sanjeev",
  booktitle = "Proc. ICASSP",
  pages = "12006--12010",
  year = 2024
}

@Inproceedings{desplanques20_interspeech,
  title = "{ECAPA-TDNN}: Emphasized Channel Attention, Propagation and Aggregation in {TDNN} Based Speaker Verification",
  author = "Brecht Desplanques and Jenthe Thienpondt and Kris Demuynck",
  booktitle = "Proc. Interspeech",
  pages = "3830--3834",
  year = 2020
}

@InProceedings{deng2019arcface,
  title = "{ArcFace}: Additive Angular Margin Loss for Deep Face Recognition",
  author = "Deng, Jiankang and Guo, Jia and Xue, Niannan and Zafeiriou, Stefanos",
  booktitle = "Proc. CVPR",
  year = 2019
}

@Article{zhao2021speakin,
  title = "The {SpeakIn} system for {VoxCeleb} speaker recognition challenge 2021",
  author = "Zhao, Miao and Ma, Yufeng and Liu, Min and Xu, Minqiang",
  journal = "arXiv preprint arXiv:2109.01989",
  year = 2021
}

@Article{jung2024espnet,
  title = "{ESPnet-SPK}: Full pipeline speaker embedding toolkit with reproducible recipes, self-supervised front-ends, and off-the-shelf models",
  author = "Jung, Jee-weon and Zhang, Wangyou and Shi, Jiatong and Aldeneh, Zakaria and Higuchi, Takuya and Theobald, Barry-John and Abdelaziz, Ahmed Hussen and Watanabe, Shinji",
  journal = "arXiv preprint arXiv:2401.17230",
  year = 2024
}

@InProceedings{wang2025geolocation,
  title = "Geolocation-Aware Robust Spoken Language Identification",
  author = "Wang, Qingzheng and Shim, Hye-jin and Sun, Jiancheng and Watanabe, Shinji",
  booktitle = "Proc. IEEE ASRU",
  year = 2025
}

@Book{myers_scotton93,
  title = "Duelling Languages: Grammatical Structure in Codeswitching",
  author = "Myers-Scotton, C.",
  publisher = "Clarendon Press",
  year = 1993
}

@Inproceedings{peng2023prompting,
  title = "Prompting the Hidden Talent of Web-Scale Speech Models for Zero-Shot Task Generalization",
  author = "Peng, Puyuan and Yan, Brian and Watanabe, Shinji and Harwath, David",
  booktitle = "Proc. Interspeech",
  pages = "396--400",
  year = 2023
}

@Inproceedings{valk2021voxlingua107,
  title = "{VoxLingua107}: a dataset for spoken language recognition",
  author = "Valk, J{\"o}rgen and Alum{\"a}e, Tanel",
  booktitle = "Proc. IEEE SLT",
  pages = "652--658",
  year = 2021
}

@Article{burchell2024code,
  title = "Code-switched language identification is harder than you think",
  author = "Burchell, Laurie and Birch, Alexandra and Thompson, Robert P and Heafield, Kenneth",
  journal = "arXiv preprint arXiv:2402.01505",
  year = 2024
}

@Article{rangan2020exploiting,
  title = "Exploiting spectral augmentation for code-switched spoken language identification",
  author = "Rangan, Pradeep and Teki, Sundeep and Misra, Hemant",
  journal = "arXiv preprint arXiv:2010.07130",
  year = 2020
}

@Article{li2023simple,
  title = "Simple yet effective code-switching language identification with multitask pre-training and transfer learning",
  author = "Li, Shuyue Stella and Xiao, Cihan and Li, Tianjian and Odoom, Bismarck",
  journal = "arXiv preprint arXiv:2305.19759",
  year = 2023
}

@Inproceedings{jia23b_interspeech,
  title = "A Compact End-to-End Model with Local and Global Context for Spoken Language Identification",
  author = "Fei Jia and Nithin Rao Koluguri and Jagadeesh Balam and Boris Ginsburg",
  booktitle = "Proc. Interspeech",
  pages = "5321--5325",
  year = 2023
}


\end{document}